\documentclass[twocolumn, trackchanges]{aastex63}
\usepackage{xspace}

\usepackage{institutions}

\newcommand\rhoopha{$\rho$ Oph A\xspace}
\newcommand\rhoophe{$\rho$ Oph E\xspace}
\newcommand\hawcplus{HAWC{\tt+}\xspace}
\newcommand\planck{\emph{Planck}\xspace}
\newcommand\herschel{\emph{Herschel}\xspace}

\shorttitle{Relative Orientation of Magnetic Field and Elongated Cloud Structure}
\shortauthors{Lee et al.}

\begin{document}
\title{\hawcplus/SOFIA Polarimetry in {L1688}: Relative Orientation of Magnetic Field and Elongated Cloud Structure}

\correspondingauthor{Dennis Lee}
\email{dennislee@u.northwestern.edu}

\author[0000-0002-3455-1826]{Dennis Lee}
    \affiliation{\CIERA}
    \affiliation{\NU}
\author{Marc Berthoud}
    \affiliation{\CIERA}
    \affiliation{\Uchicagoengineering}
\author[0000-0002-9209-7916]{Che-Yu Chen}
    \affiliation{\UVA}
    \affiliation{\LLNL}
\author[0000-0002-5216-8062]{Erin G. Cox}
    \affiliation{\CIERA}
\author{Jacqueline A. Davidson}
    \affiliation{\UWA}
\author[0000-0002-3566-6270]{Frankie J. Encalada}
    \affiliation{\UIUrbana}
\author[0000-0002-4666-609X]{Laura M. Fissel}
    \affiliation{\Queens}
\author[0000-0003-2118-4999]{Rachel Harrison}
    \affiliation{\UIUrbana}
\author[0000-0003-4022-4132]{Woojin Kwon}
    \affiliation{\SeoulNationalUniversity}
    \affiliation{\SNUAstroResearchCenter}
\author{Di Li}
    \affiliation{\ChinaNAO}
    \affiliation{\ChineseAcademy}
\author{Zhi-Yun Li}
    \affiliation{\UVA}
\author[0000-0002-4540-6587]{Leslie W. Looney}
    \affiliation{\UIUrbana}
\author[0000-0003-1288-2656]{Giles Novak}
    \affiliation{\CIERA}
    \affiliation{\NU}
\author[0000-0001-7474-6874]{Sarah Sadavoy}
    \affiliation{\Queens}
\author[0000-0002-9650-3619]{Fabio P. Santos}
    \affiliation{\MaxPlanckHeidelberg}
\author{Dominique Segura-Cox}
    \affiliation{\MaxPlanckGarching}
\author[0000-0003-3017-4418]{Ian Stephens}
    \affiliation{\Worcester}
    
\begin{abstract}
We present a study of the relative orientation between the magnetic field and elongated cloud structures for the \rhoopha and \rhoophe regions in {L1688} {in the Ophiuchus molecular cloud}. Combining inferred magnetic field {orientation} from \hawcplus 154 $\mu$m observations of polarized thermal emission with column density maps created using \herschel submillimeter observations, we find consistent perpendicular relative alignment at scales of $0.02$ pc ($33.6\arcsec$ at $d \approx 137$ pc) using the histogram of relative orientations (HRO) technique. This supports the conclusions of previous work using \planck~{polarimetry} and extends the results to higher column densities. Combining this {\hawcplus} HRO analysis with {a new \planck HRO analysis of L1688}, the transition from parallel to perpendicular alignment {in L1688} is observed to occur at a molecular hydrogen column density of {approximately $10^{21.7}$ cm$^{-2}$}. {This value for the alignment transition column density agrees well with values found for nearby clouds via previous studies using only \planck observations.} Using {existing} turbulent, {magnetohydrodynamic} simulations of molecular clouds formed by colliding flows as a model for {L1688}, we conclude that the molecular hydrogen volume density associated with this transition is approximately {$\sim10^{4}$ cm$^{-3}$}. 
{We discuss the limitations of our analysis, including incomplete sampling of the dense regions in L1688 by \hawcplus.}
\end{abstract}


\section{Introduction}\label{sec:intro}
The interstellar magnetic field is believed to play important roles in star formation. For example, the field is thought to be one of the key factors---along with gravity and turbulence---that determines and regulates the rate at which the molecular cloud evolves to form clumps, filaments, cores, and, finally, newborn stars~\citep{2007ARA&A..45..565M}. On smaller scales, the field may also strongly influence the formation of protoplanetary disks~\citep{2014prpl.conf..173L}.

A common method for observing magnetic fields in molecular clouds is measuring the linearly polarized thermal radiation emitted by dust grains within these clouds~\citep{2000PASP..112.1215H}. While the exact physical process by which these grains are aligned remains an open question, they are generally understood to orient themselves with the long axes perpendicular to the {orientation} of the local magnetic field lines~\citep{2015ARA&A..53..501A}. As a result, the far-infrared to submillimeter thermal emission is linearly polarized perpendicular to the magnetic field, thus indirectly tracing projection of the field on the sky. 

An effective method {for} using these polarization measurements to understand the magnetic field's role is to compare the inferred {orientation} and morphology of the magnetic field to the orientations of elongated molecular cloud structures~{\citep[e.g.,][]{2009MNRAS.399.1681T, 2011ApJ...734...63S, 2013A&A...550A..38P, 2013MNRAS.436.3707L, 2013ApJ...774..128S, 2014ApJ...784..116M, 2016A&A...586A.135P, 2016A&A...586A.138P}}. \cite{2013ApJ...774..128S} introduced a statistical technique to do so known as the histogram of relative orientations (HRO) method.
When applied to synthetic polarization measurements from magnetohydrodynamic (MHD) simulations, it was found that the gas structures in column density maps showed preferential alignment with magnetic field {orientations} depending on physical conditions. {At} low column densities, gas structures showed preferred parallel alignment. With sufficiently high magnetization, this preference changes from parallel to perpendicular at higher {column} densities. 
{Notably, \cite{2013ApJ...774..128S}~has shown that super-Alfv\'enic models do not predict a transition to perpendicular alignment at any column densities below $N_{\text{H}_{2}} \approx10^{22.5}$ cm$^{-2}$ ($N_{\text{H}} \approx10^{22.8}$ cm$^{-2}$).}

{Subsequent theoretical work continued to investigate the nature of the relative orientation relationship between the magnetic field and column density structure \citep[e.g,][]{2016ApJ...829...84C, 2017A&A...607A...2S, 2020MNRAS.499.4785K, 2020MNRAS.497.4196S}. It has been suggested that HROs can be used as a tool to estimate both the volume density and the magnetic field strength at the transition \citep{2016ApJ...829...84C}.
Recent HRO analyses of numerical simulations confirmed that the transition is only clearly evident in simulations with high magnetization and that the transition density threshold is primarily dependent on the magnetization of the gas \citep[e.g.,][]{2017A&A...607A...2S, 2020MNRAS.499.4785K}. \citet{2020MNRAS.497.4196S} finds that projection effects may prevent the observation of a transition, even when one exists.
}

Application of the HRO analysis to actual molecular clouds requires {a large} number of polarization measurements. In \cite{2016A&A...586A.138P}, this method was applied to ten molecular clouds using $10\arcmin$ resolution \planck polarization maps at 353 GHz. {Most clouds (eight of ten) exhibited a transition from parallel to perpendicular alignment, with an average transition column density of $N_{\text{H}_{2}} \approx10^{21.4}$ cm$^{-2}$ ($N_{\text{H}} \approx10^{21.7}$ cm$^{-2}$). This result implies that most of the sampled clouds possess at least a moderately strong magnetic field (i.e., the clouds are either trans- or sub-Alfv\'enic).} \cite{2017A&A...603A..64S} applied the HRO method to submillimeter polarization measurements of Vela C from the BLASTPol balloon-borne polarimeter~\citep{2016ApJ...824..134F}. {A similar result was found}. 

However, the $10\arcmin$ ($\sim 0.5$ pc for clouds at {$d \sim 137$} pc) \planck resolution can only probe the cloud-scale polarization, while the magnetic field structure at sub-pc scales and higher column densities could have a more direct impact on the star forming process. {On the other hand, ALMA is capable of high sensitivity polarization measurements of cores and envelopes, but is unable to map larger scales.} Mounted on the Stratospheric Observatory for Infrared Astronomy (SOFIA), the \hawcplus far-infrared polarimeter~\citep{2018JAI.....740008H} {provides the} ability to produce higher resolution maps coupled with large detector area {at these intermediate scales.} \citep[e.g,][]{2019ApJ...872..187C, 2019ApJ...882..113S}. {This} allows us to further extend analyses done with the HRO technique.

Here, we apply the HRO method to polarimetric observations of the {L1688 region in the} Ophiuchus molecular cloud. At a distance of approximately {$137$ pc}, Ophiuchus is one of the closest star-forming clouds~\citep{2008hsf2.book..351W,  2019ApJ...879..125Z}. {In order to investigate a range of {column} densities, we use \planck { polarization measurements} for the low {column} densities and \hawcplus observations of the \rhoopha and \rhoophe regions within L1688 for the high {column} densities.} Studies of L1688 have revealed a large and varied population of protostars, making it an attractive target for studying low-mass star formation~{\citep{1998A&A...336..150M, 2009ApJ...692..973E, 2019ApJS..245....2S}}. 

{The} paper is organized as follows: Section~\ref{sec:observations} describes our \hawcplus observations and presents \herschel column density maps {and \planck~{Stokes parameter measurements}} of L1688. Section~\ref{sec:methods} summarizes the HRO method and its application here to the \hawcplus~{and \planck data}. In Section~\ref{sec:results}, we present our main results. Section~\ref{sec:discussion} discusses the results and their implications{.} Finally, Section~\ref{sec:conclusion} provides a summary.

\section{Observations}\label{sec:observations}

    \begin{figure*}
        \includegraphics[width=\textwidth]{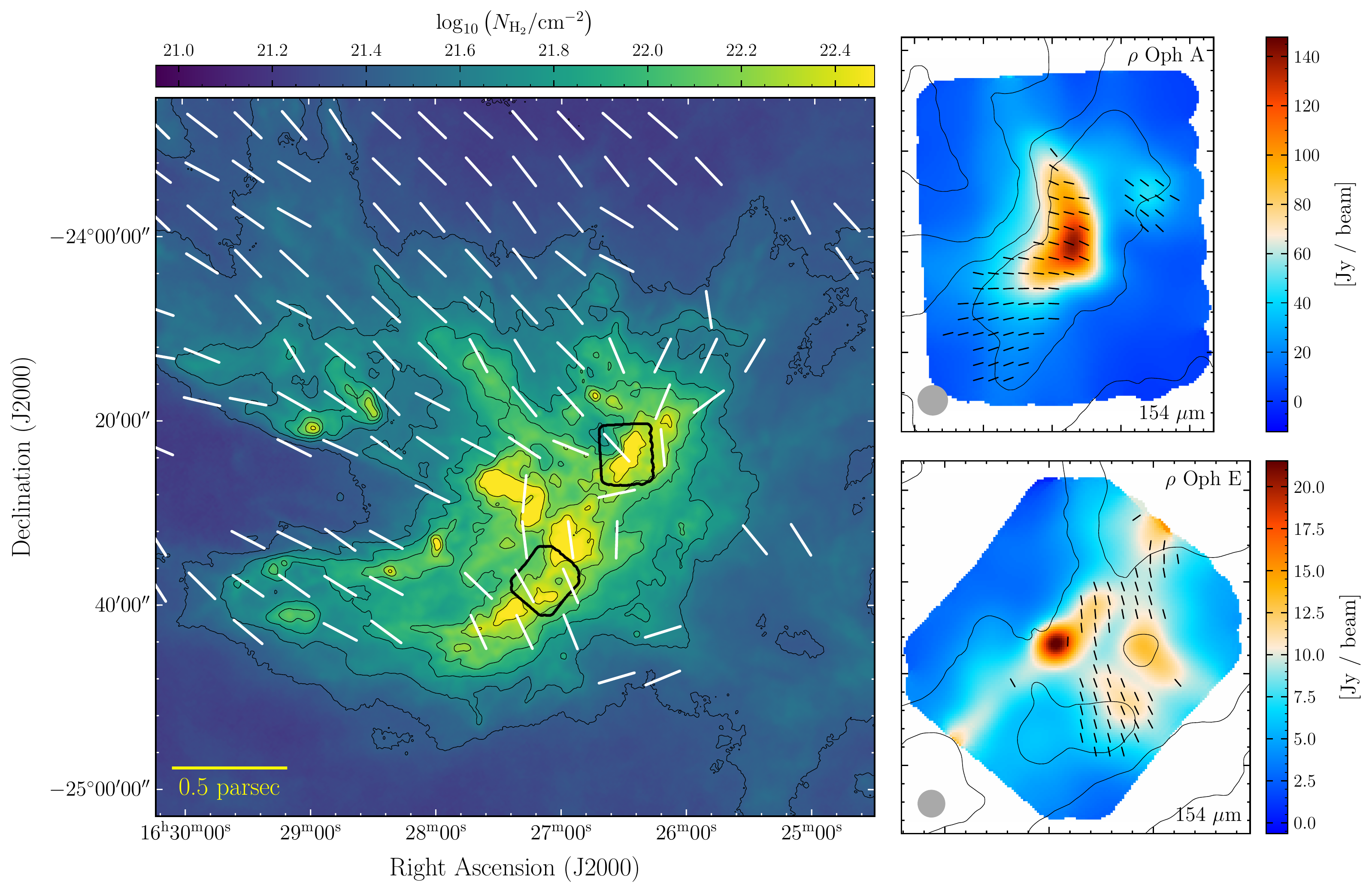}
        \caption{\emph{Left:} \herschel column density map of L1688 shown in contours and colormap. Black outlines indicate the regions of \rhoopha and \rhoophe observed using \hawcplus. {White line segments represent the native resolution inferred magnetic field orientation as observed by \planck at 353 GHz (Section \ref{subsec:obs:planck}). Line segments are shown where {$P / \sigma_P > 3$}, with spacing corresponding to the \planck beamsize ($\sim 5 \arcmin$){, and are drawn with uniform length.}} {A scalebar is shown at the lower~{left} of the panel.}
        \emph{Right:} \hawcplus 154 $\mu$m total intensity {is shown as a colormap for} \rhoopha (top) and \rhoophe (bottom) smoothed to $36.3\arcsec$ resolution to match the Herschel column density resolution of our HRO analysis {(see Sections~\ref{subsec:obs:herschel} and \ref{subsec:methods:hro})}. {The magnetic field {orientation}, also at $36.3\arcsec$ resolution, is indicated by the Nyquist-sampled uniform length black line segments}. These are limited to measurements meeting the criteria described in {Section~\ref{subsec:obs:hawc}} (e.g., $p/\sigma_p > 3$). The beam is depicted in the lower left corners. For reference, the \herschel column density is indicated by the black contours. Contour levels {for all \herschel maps are in units of} $\log_{10} \left(N_{\text{H}_2} / \text{cm}^{-2} \right)$, {with the highest contour at 22.4 and decreasing in steps of 0.2.} \label{triple-figure}}
    \end{figure*} 
    
    \begin{figure*}
        \includegraphics[width=\textwidth]{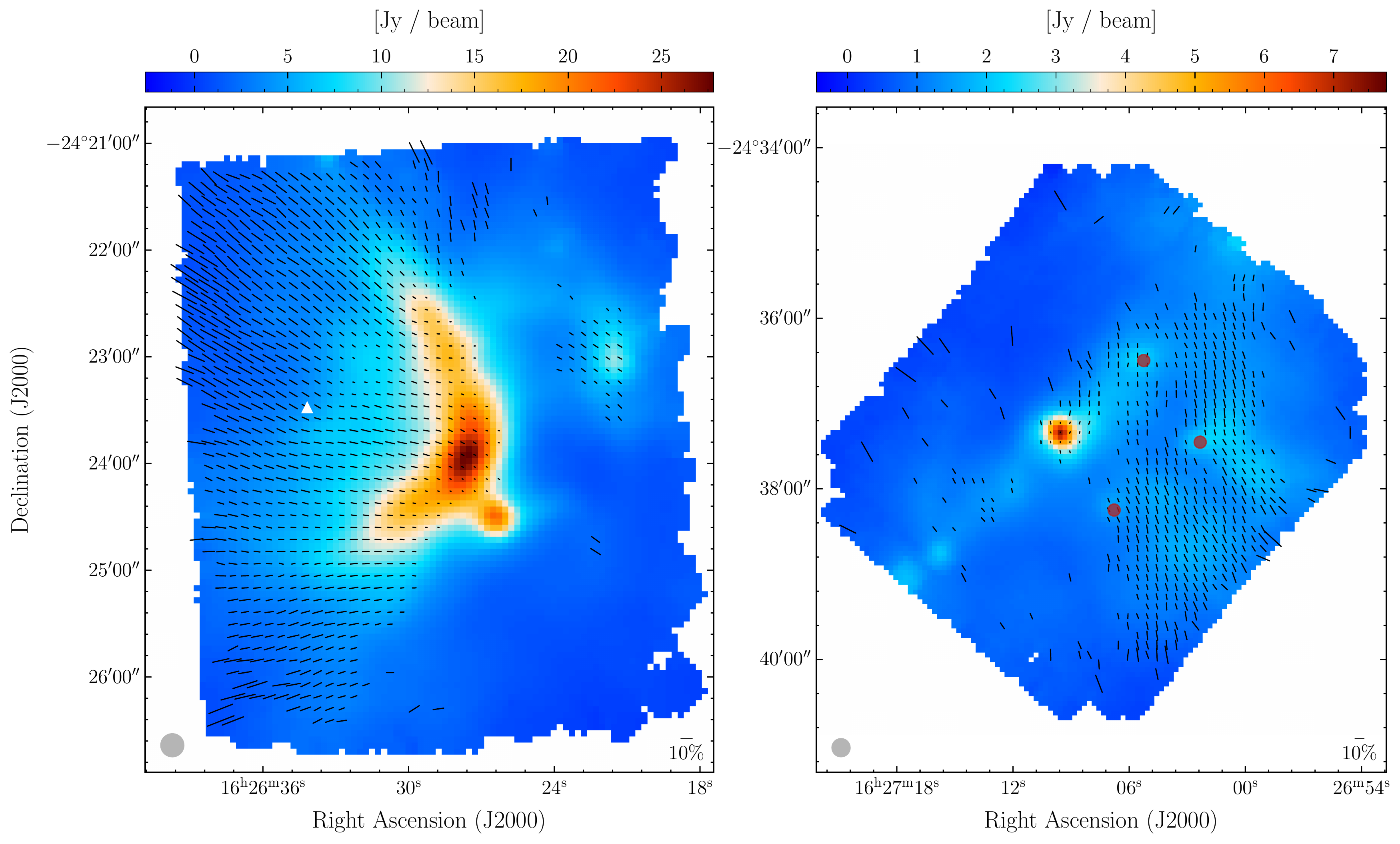}
        \caption{\hawcplus 154 $\micron$ total intensity maps and inferred magnetic field {orientation} for \rhoopha (left) and \rhoophe (right). The inferred field {orientation} is indicated by the Nyquist-sampled black line segments. These line segments are scaled based on the polarization percentage (note the 10\% scalebar in the lower right of each map) and are limited to measurements meeting the criteria described in {Section~\ref{subsec:obs:hawc}} (e.g., $p/\sigma_p > 3$). In the left figure, the white triangle marks the location of Oph S1~\citep{1988ApJ...335..940A, 2003PASJ...55..981H}. In the right figure, the red circles indicate the locations of three protostars. These are in addition to Elias 29, a Class I protostar located at the source peak. Clockwise from the source peak{, the three protostars} are LFAM 26, WL 16, and WL 17, all believed to be Class I protostars~\citep{2009ApJ...692..973E}. The beam size {of $13.6\arcsec$} is shown in the lower left of each figure. \label{observables}}
    \end{figure*}

\subsection{\hawcplus/SOFIA Observations}\label{subsec:obs:hawc}
Observations of \rhoopha and \rhoophe were made using the \hawcplus polarimeter mounted on the Stratospheric Observatory for Infrared Astronomy (SOFIA). Both sources were observed in Band D ($154\ \micron$) providing a full width at half maximum (FWHM) resolution of $13.6 \arcsec$. \rhoopha was observed in May of 2017 as part of the \hawcplus Guaranteed Time Observing (GTO) program. \rhoophe was observed on July 7th, 2018 (AOR: 06\textunderscore0116\textunderscore6). 

Observations of \rhoophe were made in the matched-nod-chop mode with a chopping frequency of 10.2 Hz and a chop throw of $430 \arcsec$. Measuring from equatorial North and increasing toward the East, we used a chop angle of $60\degr$. The observations of \rhoophe totaled 746 seconds on-source. The observations were divided into seven `dither sets' with each set consisting of four independent pointings. The dither offset between each independent pointing was $40\arcsec$ with each independent pointing {containing observations at four} different half-wave plate {angles}.

Observations of \rhoopha were also made in the matched-nod-chop mode and totaled 833 seconds. Further details on the observations of \rhoopha can be found in \cite{2019ApJ...882..113S} where the data were first presented and discussed. The regions observed of \rhoopha and \rhoophe are indicated on the \herschel-derived column density map of L1688 by solid black outlines in Figure~\ref{triple-figure} {\citep{2010A&A...518L.102A, 2020A&A...638A..74L}}. 

The observations for both regions were manually reduced using the \hawcplus Data Reduction Pipeline following the procedure described by \cite{2019ApJ...882..113S}. The process is briefly summarized here. 

{The raw time ordered data are first demodulated to account for the chopping. During this process, we also discard data affected by erroneous telescope movement or other data acquisition errors.} Flat-fielding is then done to calibrate for any pixel to pixel gain variations. {Data from dead or noisy pixels are then removed. Next, the demodulated time ordered data are combined into four sky images per independent pointing, i.e., one image per half-wave plate position. These four images are then summed and differenced to obtain final {Stokes \textit{I}, \textit{Q}, and \textit{U}} maps for each independent pointing.} After flux calibrations and corrections for atmospheric opacity are completed, the results from each {independent pointing} are combined to create final {Stokes \textit{I}, \textit{Q}, and \textit{U} maps}. The final polarization percentage is also computed and debiased. For the {total} flux calibration, we estimate an absolute uncertainty of $20 \%$. 
The $\chi^2$ statistic is then computed for the entire data set in order to evaluate the consistency between each {repeated measurements}. The purpose is to test for additional sources of uncertainties~\citep{2011ApJ...732...97D, 2013ApJ...770..151C}. The typical {cause of} these additional uncertainties is noise that is correlated across instrument pixels. The final reported uncertainties in {Stokes \textit{I}, \textit{Q}, and \textit{U}} {are} then inflated to adjust for {the} underestimation.

Next, we reject polarization measurements based on several different criteria. Measurements possessing a degree of polarization less than three times the corresponding polarization uncertainty ($p < 3\sigma_p$) or polarization angle uncertainty greater than 10 degrees {are} rejected. Nonphysical measurements of polarization ($p > 50\%$) {are} also rejected. {The cause of these high $p$ values is not known, but less than 0.5\% of the data were removed as a result of this cut.} Any polarization measurement with corresponding total intensity less than $1\%$ of the measured peak ($I < 0.01 \ \text{peak}(I)$) or total intensity uncertainty greater than $10\%$ of the total intensity (i.e., $S/N < 10$) {was} excluded. 

Since both \rhoopha and \rhoophe were observed in the matched-nod-chop mode, we also consider contamination from polarization in the reference beam areas. To do so, we {flag} sky positions where the differenced polarized flux is less than the average polarized flux for the  two corresponding reference beam locations. This method is based on the analysis described in \cite{1997ApJ...487..320N} and \cite{2019ApJ...872..187C}. For each sky pixel, where $p_m$ is the measured polarization fraction and $p_r$ is the polarization in the reference beam, we reject pixels where:
\begin{equation}\label{eq:refbeamcut}
    p_m^2 < \left(p_r w \right)^2
\end{equation}
Here, $w$ is the expected ratio of reference beam intensity to differenced intensity,
\begin{equation}\label{eq:refbeamratio}
    w = \left( \frac{\bar{I_r}}{I_s - \bar{I_r}} \right)
\end{equation}
{where $I_s$ is the intensity in the source region. \herschel 160 $\mu$m intensity maps of the region are used to estimate $w$. As each source was observed with two reference beam areas, symmetrically located on either side of the central area, $\bar{I_r}$ represents the average of the intensities in these two areas.} We use a conservative estimate where the reference beam area has a uniform polarization of $p_r = 0.1$. 

Any pixels that are flagged by the condition described in Equation~\ref{eq:refbeamcut} are discarded as these are considered too contaminated to be of use. For measurements that survive, there nonetheless remains the possibility of some contamination, even if to a lesser degree. A more detailed analysis designed to more carefully quantify the impact of the reference beam---specifically on the measured polarization angle and its effect on subsequent analyses---{can be found in Section~\ref{subsec:results:transition} and Appendix~\ref{app:uncertainties}}.

The resulting \hawcplus observations of \rhoopha and \rhoophe are shown in Figure~\ref{observables}{, where the} magnetic field {orientation} is inferred by rotating the polarization measurements by 90\degr.

Comparing the resulting reduction of \rhoopha presented here {with} the one presented in  \cite{2019ApJ...882..113S}, the basic morphology of the magnetic field is generally the same. However, owing to the different approaches used to consider the reference beam effects, there are small variations in the spatial coverage of the polarization measurements. Beyond this, we note several other minor {differences}. First, the pointing is verified based on comparison with the \herschel~{PACS} 160 $\mu$m maps of Ophiuchus {obtained from the \herschel Science Archive} {\citep{2010A&A...518L...2P}} and thus differs slightly between the two results. Additionally, the $\chi^2$ analysis finds an underestimation of the uncertainties by approximately $36\%$ versus $38\%$ in \cite{2019ApJ...882..113S}.

\subsection{\planck Stokes Parameters Maps}\label{subsec:obs:planck}
{In order to investigate the low {column} density regions of L1688, we use all-sky linear polarization measurements from \planck. At 353 GHz, these measurements have a native resolution of $5\arcmin$. We use {Stokes \textit{Q} and \textit{U}} maps from Data Release 3 (DR3) available on the {\planck} Legacy Archive \citep{2020A&A...641A...1P}. The tangent plane projection is produced for the L1688 region, defined by a $1.3^\circ \times 1.3^\circ$ region (see Figure \ref{triple-figure}). To remove low quality measurements, polarization measurements possessing a polarized intensity less than three times the corresponding polarized intensity uncertainty ($P < 3\sigma_P$) are rejected. The resulting inferred magnetic field orientations are plotted in Figure \ref{triple-figure}. Compared to the smoothed $10\arcmin$ resolution 353 GHz \planck~{inferred magnetic field} used in \citet{2016A&A...586A.138P}, {these} higher resolution data {are} largely consistent. } 

\subsection{\herschel Column Density Maps}\label{subsec:obs:herschel}
{A column density map for L1688 was obtained from the publicly available \herschel Gould Belt Survey (HGBS) Archive}~\citep{2010A&A...518L.102A}. The 160, 250, 350, and 500 $\micron$ \herschel observations of Ophiuchus were fit to a modified blackbody function in order to produce {\herschel column density maps} at a resolution of $36.3\arcsec$ (See  \citealt{2020A&A...638A..74L} for more details). {The column density map is shown in Figure~\ref{triple-figure}.} In addition to the $36.3\arcsec$ resolution column density map, a {``high-resolution''} column density map with an effective resolution of $18.2\arcsec$ created through a multi-scale decomposition method is also available~\citep{2013A&A...550A..38P}. However, to be conservative, we have opted to use the native $36.3\arcsec$ resolution column density map. 

\section{Histogram of Relative Orientations}\label{sec:methods}
{In Figure \ref{triple-figure}, we see that the inferred magnetic field orientation is preserved across the \planck and \hawcplus measurements despite the factor of nine difference in angular resolution. This continuity has motivated us to conduct a joint \hawcplus/\planck histogram of relative orientations (HRO) analysis of L1688.} The {HRO} method is designed to characterize the orientation of the magnetic field in the context of column density structures~\citep{2013ApJ...774..128S, 2016A&A...586A.138P,2017ApJ...842L...9H, 2019ApJ...878..110F}. In particular, this method describes any preference for parallel or perpendicular alignment of the magnetic field with respect to elongated structures seen in column density maps. {In the present paper, we apply this method to both the \hawcplus and \planck~{datasets} described in Section \ref{sec:observations}. The procedure for each case is nearly identical. Differences are highlighted in the description of the analysis below.}

\subsection{HRO Construction}\label{subsec:methods:hro}
The relative orientation between the density structure and the magnetic field {as projected onto} the plane of the sky can be characterized by the angle $\phi$, defined as the relative angle between the projected magnetic field vector $\hat{B}$ and the line tangent to the local iso-contour of gas column density (Note that this is as applied to observations in 2D; section~\ref{subsec:voltrans} will discuss {the} analogous application in 3D). Equivalently, this can be defined as the relative orientation angle between the polarization vector ($\hat{E}$) and the gradient of the column density structure ($\nabla N$). {The HRO is the distribution of the relative orientations $\phi$.} 

For consistency, {prior to the \hawcplus HRO analysis}, we smooth the $13.6\arcsec$ resolution \hawcplus~{Stokes \textit{Q} and \textit{U}} observations to the $36.3\arcsec$ resolution of the \herschel column density maps. {{Smoothed \hawcplus polarimetry results are shown in Figure \ref{triple-figure}.} Similarly, for the \planck analysis we smooth the $36.3\arcsec$ resolution \herschel column density maps to the $5\arcmin$ resolution of the {\planck} data.} 

{For the \hawcplus analyses, following the IAU convention, the} angle of $\hat{E}$ is computed from the {\textit{Q} and \textit{U}} values by:
\begin{equation}
    \psi_{\hat{E}} = \frac{1}{2} \arctan \left( U, Q \right)
\end{equation}
{Due to the different conventions followed by the \planck Stokes parameters, the following is used for \planck data: }
\begin{equation}
    \psi_{\hat{E}} = \frac{1}{2} \arctan \left( -U_{\text{\planck}}, Q_{\text{\planck}} \right)
\end{equation}
{Details can be found in Section 2.1 of \citet{2015A&A...576A.104P}.}

{All data is then regridded onto a 3$\arcsec$ pixel size grid.} We use a Gaussian derivative kernel to calculate the gradient ($\nabla N$). {For the column density maps to be compared with the \hawcplus~{polarization angle}, we chose a kernel with a FWHM of $12 \arcsec$. For the column density maps to be compared with the {\planck polarization angle}, we chose a kernel with a FWHM of $30 \arcsec$.} 
{These kernel sizes are} chosen such that they are large enough to smooth out and remove any potential edge or corner effects that may create erroneous gradient vectors, {and yet} small enough such that {no significant degradation of} the resolution of our gradient map {occurs}.
    
Using both the polarization vector and the gradient vector, the relative orientation angle ($\phi$) is then computed using:
\begin{equation}
    \phi = \text{arctan}(|\nabla N \times \hat{E}|, \nabla N \cdot \hat{E})
\end{equation}
Under this convention, $\phi = 0\degr$ represents {parallelism between} magnetic {field} and elongated column density structures{, while} $\phi = \pm 90\degr$ represents {perpendicularity between} magnetic {field} and elongated column density structures. 

{Next, for both the \planck and \hawcplus $\phi$ maps separately, the respective map is divided into four bins that are ordered by $N_{\text{H}_2}$. The column density ranges for the four bins are determined by constraining each bin to have the same number of data points. For each of these four bins, an HRO is then produced. This results in a total of eight HRO analyses, four for \planck and four for \hawcplus.}

{Due to the fact that the \hawcplus~{observations} covers only \rhoopha and \rhoophe, {they sample} only a minor portion of the high column density sightlines in L1688. On the other hand, the \planck~{maps covers} the majority of the low column density sky area in the region studied. Accordingly, we make a cut on the column density for the \hawcplus data to remove column densities satisfying $\log_{10}\left(N_{\text{H}_2}\right) < 22.3$. At column densities lower than this threshold, the \hawcplus observations only sample $\sim1\%$ of the available sky area in L1688, whereas for column densities above this threshold, the undersampling is not as severe. The effects of this ``sampling uncertainty'' on our conclusions is discussed in Sections \ref{subsec:results:transition} and \ref{subsec:voltrans}. Note that the column density cut is applied before the \hawcplus data are divided into four column density bins (see previous paragraph).}

\subsection{Relative Orientation Parameter}\label{subsec:methods:param}
The characteristic shape of the histogram describing the distribution of $\phi$ can be represented using $\xi$,~{the normalized version of the the HRO shape parameter \citep{2013ApJ...774..128S, 2016A&A...586A.135P}}. The parameter is defined as:
\begin{equation}
    \xi = \frac{A_c - A_e}{A_c + A_e}
\end{equation}
$\xi$ compares the area of the center of the histogram ($A_c$) with the area of the extremes of the histogram ($A_e$). The central area ($A_c$) is defined as the area of the region: $-22.5^\circ < \phi < 22.5^\circ$. The extremes area ($A_e$), is defined as the union of the two regions: $-90^\circ < \phi < -67.5^\circ$ and $67.5^\circ < \phi < 90^\circ$. 
    
Using this definition we can see that $\xi$ holds a value between $-1$ and $1$. $\xi > 0$ indicates a histogram with a peak between $-22.5^\circ$ and $22.5^\circ$. This corresponds to a preference for parallel alignment between the elongation of the gas structure and the magnetic field. Conversely, when $\xi < 0$, this indicates a histogram where the magnetic field is largely perpendicular to the column density contours. In a situation of no alignment preference, the corresponding histogram will be largely flat, and the shape parameter will be approximately zero ($\xi \approx 0$).

The uncertainty in $\xi$ {is} $\sigma_\xi$ defined as:
\begin{equation}
    \sigma_\xi^2 = \frac{4\left(A_e^2\sigma_{A_c}^2 + A_c^2\sigma_{A_e}^2\right)}{\left(A_c + A_e\right)^4}
\end{equation}
where $\sigma_{A_e}^2$ and $\sigma_{A_c}^2$ are the variances of $A_e$ and $A_c$. As explained in \cite{2016A&A...586A.138P}, this uncertainty represents the `jitter' of the histogram. 

\begin{figure}
        \includegraphics[width=\columnwidth]{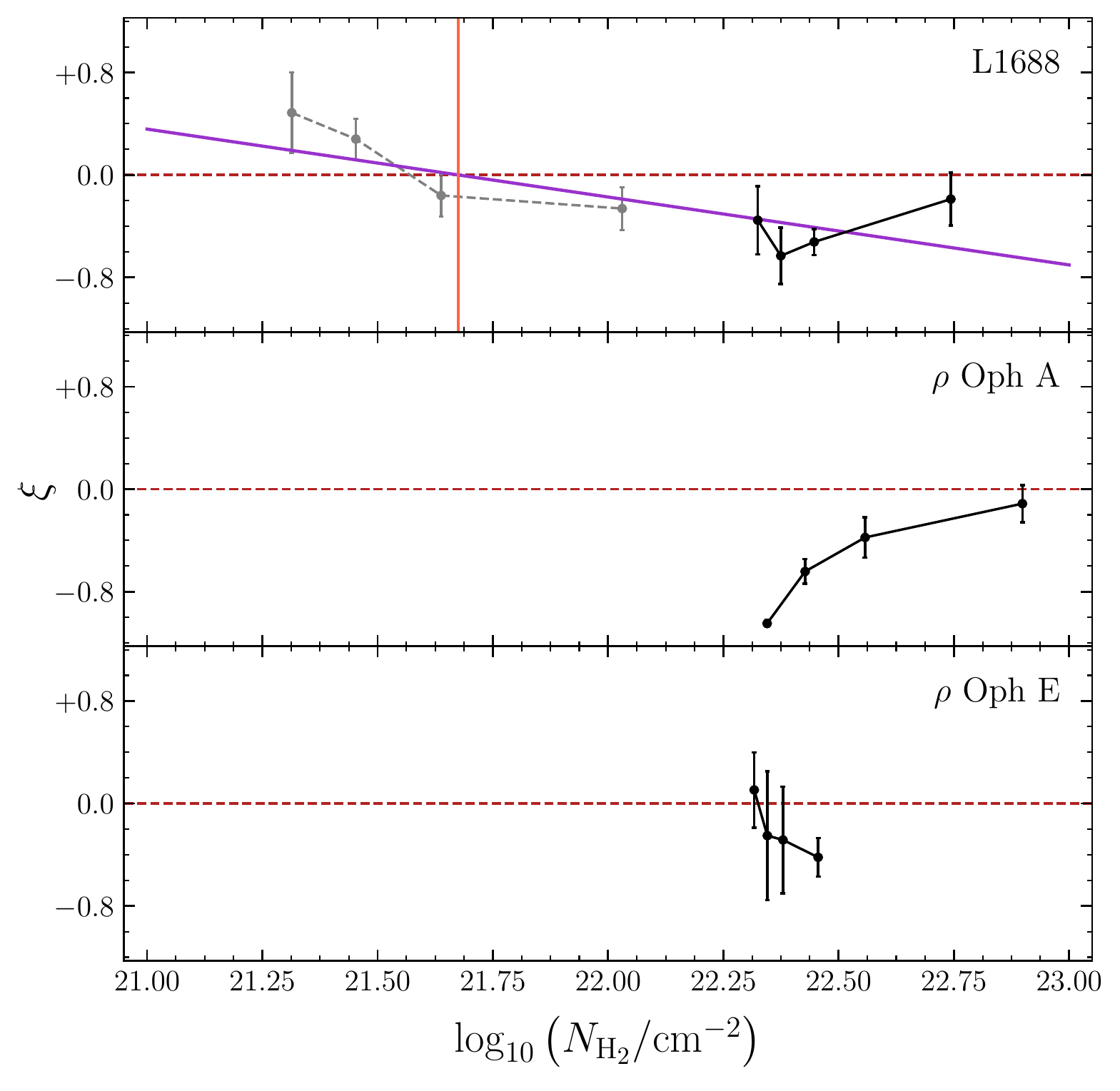}
        \caption{The HRO shape parameter as a function of column density for {L1688 (top), $\rho$ Oph A (middle), and $\rho$ Oph E (bottom).} {Solid black and dashed grey lines indicate \hawcplus and \planck ~{data}, respectively.} All errorbars indicate $1\sigma$ uncertainties.  {In the top panel, the resulting least-squares fit of the merged \planck and \hawcplus data is shown in purple. The estimated transition column density ($10^{21.7}$ cm$^{-2}$) is indicated by the vertical red line.} \label{hro_result}}
    \end{figure}

{The HRO shape parameter ($\xi$) is computed for each of the four bins consisting of \planck data and also for each of the four bins consisting of \hawcplus data. We then plot $\xi$ as a function of column density bin (specifically of the median density value for the bin). This is shown in the top panel of Figure~\ref{hro_result}. Also shown in Figure~\ref{hro_result} are \hawcplus HRO results for \rhoopha and \rhoophe separately (lower panels).}

\section{Results}\label{sec:results} 

From Figure~\ref{hro_result}, we see that for all three calculations of the $\xi$ values from \hawcplus ({L1688, \rhoopha, \rhoophe}), we obtain {predominantly} negative values indicative of perpendicular alignment {(black points). As discussed in Section \ref{sec:intro}, the HRO analysis in \cite{2016A&A...586A.138P} noted a transition from parallel alignment at low to perpendicular alignment at high {column} densities (i.e., from positive $\xi$ to negative $\xi$) in molecular clouds. This transition, as observed by \planck in L1688, can be seen in the grey points of Figure 3.}

Viewing the {combined \planck/\hawcplus} result {presented here} as an extension of the \cite{2016A&A...586A.138P} analysis, we find evidence that the trend initially observed in \cite{2016A&A...586A.138P} continues {smoothly} to higher {column} density regimes{, at least for L1688} {(see Section \ref{subsec:results:transition})}. 

\section{Discussion}\label{sec:discussion}

\subsection{Alignment of Magnetic Field and Elongated Cloud Structures}\label{subsec:paravperp}
The parallel \text{vs.} perpendicular magnetic field behavior has been theoretically predicted in models of star formation~{\citep{2013ApJ...774..128S, 2014prpl.conf...27A}.} In the model envisioned by \cite{2014prpl.conf...27A}, filamentary structures play a major role in the star formation process. These molecular filaments are elongated structures that may be produced via  mechanisms such as converging flows. They appear to accumulate mass via relatively lower density `striations' running perpendicular to the axis of the relatively higher density filament. In this picture, these low density striations are expected to be parallel to field lines. In turn, the field lines are thus perpendicular to the major axis of the higher density filament. This would result in the low density-parallel to high density-perpendicular behavior which is consistent with what {we observe in L1688 (see Figure \ref{hro_result})}. 

{This behavior is also supported by a number of prior observational studies. As discussed in Section \ref{sec:intro}, \citet{2016A&A...586A.138P} observed a transition from parallel to perpendicular alignment with increasing column density in ten giant molecular clouds. This was also observed in Vela C \citep{2017A&A...603A..64S}. \citet{2019A&A...629A..96S} combined observations of large scale (\planck) magnetic field and small scale (\herschel) column density to understand the orientation between column density structures and the magnetic field. While distinctly different from what is done here, their study also noted a trend of column density \text{vs.} magnetic field relative orientation changing from preferentially parallel to perpendicular with increasing column density. Recent improvements in polarimetric instrumentation on the James Clerk Maxwell Telescope (JCMT) have enabled magnetic field observations at smaller scales and higher column densities than are accessible via \planck data~\citep{2017ApJ...842...66W}. Though not using the HRO method, recent studies using the JCMT~\citep[e.g.,][]{2017ApJ...842...66W, 2018ApJ...859..151L, 2019ApJ...883...95S, 2020ApJ...899...28D, 2021ApJ...907...88P} have generally observed perpendicular alignment at these smaller scales.}

Observations of Serpens South using \hawcplus as well as archival near-infrared data also noted a general perpendicular behaviour between the magnetic field and elongated structures~\citep{2020NatAs...4.1195P}. However, one of the filaments exhibited a transition from perpendicular back to parallel alignment at higher column densities ($N_{\text{H}_2} \approx 10^{22.3}$  cm$^{-2}$) suggesting that magnetic alignment behavior may be more complex. This return to parallel alignment is hinted at when examining our analysis of \rhoopha in Figure~\ref{hro_result}. Given the uncertainties however, observation of this putative new transition in \rhoopha is {tentative}. 

\subsection{Estimating the Transition Column Density}\label{subsec:results:transition}

{As can be seen in Figure \ref{hro_result}, our joint analysis of L1688 spans a sufficiently large range of column densities to include both the low-density/parallel and high-density/perpendicular regimes.}
{By fitting} the simple linear trend line described in \cite{2016A&A...586A.138P} and \cite{2017A&A...603A..64S}, we can estimate a value for the transition column density. Using the Levenberg-Marquardt least-squares optimization algorithm as implemented in the Python \texttt{scipy} library, we find {$N_{\text{H}_2, \text{tr}} \approx 10^{21.7}$ cm$^{-2}$}. The result of the fit is shown in Figure~\ref{hro_result}. The statistical uncertainty from the fit is a factor of $\approx 1.25$. Note that this uncertainty {is obtained by} accounting for the `jitter' of the histogram ($\sigma_\xi$; indicated in Figure~\ref{hro_result} by the errorbars). 
{Note also that prior to fitting the data, we set the uncertainty for a given value of $\xi$ equal to the greater of (a) its corresponding uncertainty $\sigma_\xi$, and (b) the median value of $\sigma_\xi$ for all data points from the corresponding instrument (\planck or \hawcplus) used in the fit.  The reason for this procedure is that artificially low values of $\sigma_\xi$ can be obtained when sampling regions with high spatial correlations of the magnetic field \citep{2016A&A...586A.135P, 2016A&A...586A.138P}, and these can skew the fit.} 
While we have used a simple linear trend line to model the transition in $\xi${, the} exact form and manner by which the transition is expected to occur is unknown. This is a significant uncertainty in the determination of $N_{\text{H}_2, \text{tr}}$ and is likely several times larger than the statistical uncertainty from the fit. 

{Other sources of error in our estimate of $N_{\text{H}_2, \text{tr}}$ are: (a) the reference beam (Section \ref{subsec:obs:hawc}), (b) the combining of data from two different telescopes, and (c) the sampling uncertainty. In Appendix \ref{app:uncertainties}, we assess the first two of these problems. We conclude that given the compounding effects of all sources of uncertainty except for (c) the total uncertainty in the value of $N_{\text{H}_2, \text{tr}}$ is roughly a factor of three. However, the uncertainties associated with (c), the incomplete sampling of L1688, are difficult to quantify. While \hawcplus observations cover only the \rhoopha and \rhoophe regions, our \planck analysis pertains to {L1688} as a whole. In addition to \rhoopha and \rhoophe, L1688 also includes other dense regions (see Figure \ref{triple-figure}). We have discarded \hawcplus data for sightlines having column density below $\log_{10}\left(N_{\text{H}_2}\right) = 22.3$ (see Section \ref{subsec:methods:hro}), but even with this restriction, the remaining \hawcplus HRO sightlines sample only $\sim10\%$ of the corresponding high column density sky area in L1688.}

{A full treatment including the other higher {column} density subregions of L1688 would be valuable. Maps with other polarimeters do exist for at least one target~\citep[$\rho$ Oph C;][]{2019ApJ...877...43L}, but a proper HRO treatment of such maps would require consideration of uncertainties associated with reference beam contamination for instruments other than \hawcplus (see Appendix \ref{app:uncertainties}) which is beyond the scope of the present paper.} {In Section \ref{subsec:voltrans}, we discuss the impact of the sampling certainty on our main conclusions.}

{Super-Alfv\'enic simulations from \citet{2013ApJ...774..128S} (as discussed in Section~\ref{sec:intro}) do not exhibit a transition to perpendicular alignment at any column density below $N_{\text{H}_{2}} \approx10^{22.5}$ cm$^{-2}$. The estimated transition column density here  is $N_{\text{H}_2, \text{tr}} \approx 10^{21.7}$ cm$^{-2}$, suggesting that L1688 is either trans- or sub-Alfv\'enic, as found in previous work for 11 clouds \citep{2016A&A...586A.138P, 2017A&A...603A..64S}. Notably, one of the clouds studied in \citet{2016A&A...586A.138P} is Ophiuchus, of which L1688 is a part.}

{Although \citet{2019A&A...629A..96S} estimated a transition column density for several regions in Ophiuchus, including L1688, their analysis is distinct from the one performed here in that it compares large scale magnetic field and small scale column density. Additionally, the region analyzed as L1688 in \citet{2019A&A...629A..96S} is not identically defined as the region in this work. Despite these differences, when compared with the $N_{\text{H}_2, \text{tr}} \approx 10^{21.7}$ cm$^{-2}$ calculated here, \citet{2019A&A...629A..96S} finds a similar transition column density for L1688: $N_{\text{H}_{2}, \text{tr}} \approx10^{21.5}$ cm$^{-2}$ ($N_{\text{H}, \text{tr}} \approx10^{21.75}$ cm$^{-2}$).}

\subsection{Estimating the Transition Volume Density}\label{subsec:voltrans}
{The transition \emph{volume} density at which the alignment preference shifts from parallel to perpendicular has been suggested by simulations to be tied to the physical properties of the cloud~\citep[e.g.,][]{2013ApJ...774..128S, 2016ApJ...829...84C}. This `critical' transition volume density $\left( n_{\text{tr}} \right)$ is thus a useful diagnostic. In this section, we will make an initial estimate of $n_{\text{tr}}$ based on our HRO analysis.}

{
\citet{2010ApJ...725..466C} reported a transition {volume} density at $n_{\text{H}} \sim 300$ cm$^{-3}$ ($n_{\text{H}_2} \sim 150$ cm$^{-3}$) for the magnetic field strength (inferred from Zeeman-splitting observations) to transition from being relatively constant ($B \propto n^0$) as a function of gas density to having a power-law relationship with the gas density ($B \propto n^{2/3}$). This \emph{scaling transition density} is believed to correspond to the point where the magnetic field transitions from being capable of providing support against gravitational collapse to a situation where it is no longer capable of doing so. As pointed out by \citet{2016ApJ...829...84C} \citep[see also][]{2017A&A...607A...2S, 2019PASA...36...29A}, this scaling transition may correspond to the alignment transition that is revealed by the HROs.
} %
\citet{2016ApJ...829...84C} used \texttt{Athena} ideal MHD simulations to study the star forming process in shock-compressed regions in the molecular clouds~\citep{2014ApJ...785...69C, 2015ApJ...810..126C}. In this model of cloud formation, the gas is initially in a super-Alfv\'enic state. After the collision of the flows, a flat, dense, sub-Alfv\'enic post-shock region is created. The snapshot when the most evolved core begins to collapse is considered. Three variants of the simulations are analyzed. Each variant was initialized with a different inflow Mach number ($\mathcal{M}$ = 5, 10, and 20), but were identical otherwise. 

In a generalization of the method described in Section~\ref{sec:methods} to three dimensions, \citet{2016ApJ...829...84C} created synthetic 3D HROs from the simulations. {Three dimensional HROs are also discussed by \citealt{2013ApJ...774..128S}.} The alignment transition volume density is then determined as where the 3D HRO shape changes from concave to convex (i.e., the 3D HRO alignment parameter moves from $>0$ to $<0$). The scaling transition volume density from each simulation is also determined by finding the volume density at which the magnetic field strength changes from being constant as function of volume density to becoming highly correlated with this quantity~\citep[akin to][]{2010ApJ...725..466C}. {Importantly, the alignment transition volume density and the scaling transition volume density introduced by \citet{2010ApJ...725..466C} are noted to roughly coincide.} {\citet{2016ApJ...829...84C} attributes this coincidence to} the alignment transition {and the scaling transition} both being governed by the same physical processes. 

Moving from 3D space to 2D, a similar change in HRO shape is seen in the synthetic observations when analyzing 2D projected maps. Analysis {shows that} this change results from the increasing importance of self-gravity, directly linking it to the transition volume density (i.e., the $N_{\text{H}_2, \text{tr}}$ value is linked to the $n_{\text{H}_2, \text{tr}}$ value). In short, by estimating $N_{\text{H}_2, \text{tr}}$ observationally via the HRO method, it is possible to {determine} the {volume density} at which the gas becomes self-gravitating.

{Although} the transition column {density values} calculated for the simulated clouds in \cite{2016ApJ...829...84C} are not the same as the transition column density value we found in {L1688}, we note that these isothermal, ideal-MHD simulations are scale-free which means the column density can be rescaled~\citep[e.g., see][]{2018MNRAS.474.5122K}. Following the discussion and convention in \cite{2018MNRAS.474.5122K} and assuming $\lambda$ is the rescaling coefficient, the physical parameters of the simulations can be rescaled as $ L \rightarrow L / \lambda$, $n \rightarrow \lambda^2 n$. This implies $N \rightarrow \lambda N$. {Without repeating the simulations, we} can therefore rescale the simulations in \cite{2016ApJ...829...84C} to have the same $N_{\text{H}_2, \text{tr}}$  as the one we derived for {L1688 (from Section \ref{subsec:results:transition})}, and use the same rescaling coefficient to calculate the corresponding rescaled $n_{\text{H}_2, \text{tr}}$ from the original $n_{\text{H}_2, \text{tr}}$ reported in Table 1 of \cite{2016ApJ...829...84C}. We obtain $\lambda$ between {0.24 and 0.60}, depending on Mach number. {After rescaling, the resulting $n_{\text{H}_2, \text{tr}}$ values range between $10^{3.8}$ and $10^{4.2}$ cm$^{-3}$.} 

{Deriving $n_{\text{H}_2, \text{tr}}$ from $N_{\text{H}_2, \text{tr}}$ can be seen as a division by an effective length. When using the simulations from \citet{2016ApJ...829...84C} to obtain $n_{\text{H}_2, \text{tr}}$, this effective length is several times smaller than the plane of sky dimensions of L1688. This is consistent with the results of numerous simulation works finding that flattened slab-like structures are typical \citep[e.g.,][]{2016ApJ...833...10I, 2019MNRAS.485.4509L}. }


{The $n_{\text{H}_2, \text{tr}}$ values we find here can} be compared to an estimate of the {same quantity} for a different molecular cloud {made by} \citet{2019ApJ...878..110F}. Using molecular line density tracers, \citet{2019ApJ...878..110F} estimated the transition volume density value of Vela C to be $\sim 10^{3}$ cm$^{-3}$ with uncertainties on the level of one order of magnitude. {Their} method combined polarization maps from the BLASTPol balloon-borne polarimeter and intensity maps of various molecular line from Mopra to estimate the {projected Rayleigh statistic} (PRS), a statistic analogous to $\xi$. By then quantifying the characteristic densities corresponding to each molecular species, the point where the PRS transitions from parallel alignment to no preferred orientation or a weakly perpendicular alignment was estimated to determine the {alignment} transition volume density. {Given the order of magnitude in uncertainties of \citet{2019ApJ...878..110F} and our own sampling uncertainties, which are difficult to quantify, we conclude that there is no significant tension between our result and that of \citet{2019ApJ...878..110F}.}

{The difference between the values we obtain here using the simulations as the framework ($10^{3.8} - 10^{4.2}$ cm$^{-3}$) and values obtained in \citet{2010ApJ...725..466C} ($\sim 150$ cm$^{-3}$) is approximately a factor of $\sim100$. The factor of three uncertainty in $N_{\text{H}_2}$ discussed in Section \ref{subsec:results:transition} propagates to approximately an order of magnitude uncertainty in $n_{\text{H}_2}$ here. However, there remains the sampling uncertainty which is difficult to quantify. Specifically, it is unknown whether analysis of the unsampled subregions of L1688 would result in generally negative $\xi$ values as found for \rhoopha and \rhoophe. Should the unsampled regions of L1688 present different HRO behaviors, the resulting transition density could change significantly. Recalculation of $N_{\text{H}_2, \text{tr}}$ for \rhoopha and \rhoophe separately show a variation of, at most, a factor of a few ($\sim2.5$). Analysis of the remaining unsampled regions will be required to constrain this variation further. Nonetheless, based on subregions sampled so far, it would appear that there may be a large ($\times100$) discrepancy between $\sim 150$ cm$^{-3}$, the value of \citet{2010ApJ...725..466C}, and $\sim10^{4}$ cm$^{-3}$ determined here.}

{Recent work reexamining and extending the Bayesian models from \citet{2010ApJ...725..466C} obtained significantly higher scaling transition volume density values. Specifically, \citet{2020ApJ...890..153J} found a transition volume density value of $n_{\text{H}_2} \sim 560$ cm$^{-3}$, lessening the the discrepancy. As more Zeeman-splitting observations become available \citep[i.e.,][]{2019ApJ...884...49T}, it is likely that this value will be updated further.}

\section{Conclusions}\label{sec:conclusion}
In this paper, we have {used 154 $\mu$m polarization observations from \hawcplus and \herschel-derived column density maps to characterize} the alignment between the magnetic field and elongated column density structures of \rhoopha and \rhoophe using the HRO method. {Using this method, we found} a preference for perpendicular alignment at higher densities. This preference is observed for each region analyzed individually as well as when analyzed together.

{Combining a \planck HRO analysis of L1688 completed at lower densities and this \hawcplus result allows an HRO analysis over scales of $\sim 0.02-3.1$ pc ($33.6\arcsec-1.3\degr$ at $d \approx 137$ pc). Using the combined dataset, we estimated the transition column density at which the L1688 region of  Ophiuchus changes from parallel alignment to perpendicular alignment to be {$N_{\text{H}_2} \approx 10^{21.7}$ cm$^{-2}$}. This is consistent with the results of \citet{2016A&A...586A.138P}.}

{To explore the implications of our results, we calculate a value for the alignment transition volume density under the assumption that the ideal-MHD colliding flow models of \citet{2016ApJ...829...84C} apply to L1688. {We conclude that our value of $N_{\text{H}_2, \text{tr}}$ implies $n_{\text{H}_2} \approx 10^{4}$ cm$^{-3}$.}}

{In the model of \citet{2016ApJ...829...84C}, the alignment transition volume density approximately equals the scaling transition volume density, which has been previously measured by \citet{2010ApJ...725..466C} for an ensemble of clouds. However, the latter is observed by \citet{2010ApJ...725..466C} to be approximately two orders of magnitude smaller than the values we find for the alignment transition volume density in L1688.}

{The reexamination of Zeeman observations by \citet{2020ApJ...890..153J} presents one possible path to help reconcile these values.} {The discrepancy might be further ameliorated if a complete sampling of the high density sightlines of L1688 changes the result of the combined \planck/\hawcplus HRO analysis.  The two separate regions we observed with \hawcplus (\rhoopha and \rhoophe) give similar results, but together represent only 10\% of all the high density sightlines in L1688. Further observations of this cloud using \hawcplus/SOFIA will be valuable. Finally, it seems likely that crucial insights can be gained by using a variety of simulations to carry out the conversion of $N_{\text{H}_2, \text{tr}}$ to $n_{\text{H}_2, \text{tr}}$ and comparing the results.}

\acknowledgments
This work is based on observations made with the NASA/DLR Stratospheric Observatory for Infrared Astronomy (SOFIA). SOFIA is jointly operated by the Universities Space Research Association, Inc. (USRA), under NASA contract NAS2-97001, and the Deutsches SOFIA Institut (DSI) under DLR contract 50 OK 0901 to the University of Stuttgart. Financial support for this work was provided by NASA through awards \#SOF~06-0116 and \#SOF~07-0147 issued by USRA to Northwestern University. This research has made use of data from the Herschel Gould Belt survey (HGBS) project (\url{http://gouldbelt-herschel.cea.fr}). The HGBS is a Herschel Key Programme jointly carried out by SPIRE Specialist Astronomy Group 3 (SAG 3), scientists of several institutes in the PACS Consortium (CEA Saclay, INAF-IFSI Rome and INAF-Arcetri, KU Leuven, MPIA Heidelberg), and scientists of the Herschel Science Center (HSC). This work {makes use} of observations obtained with \planck (\url{http://www.esa.int/Planck}), an ESA science mission with instruments and contributions directly funded by ESA Member States, NASA, and Canada. W.K. was supported by the New Faculty Startup Fund from Seoul National University. C.Y.C. acknowledges support by NSF AST-1815784. Z.Y.L. is supported in part by NASA 80NSS20K0533.
\facilities{SOFIA (\hawcplus)}
\software{
    \texttt{numpy}~\citep{harris2020array},
    \texttt{scipy}~\citep{2020SciPy-NMeth},
    \texttt{matplotlib}~\citep{Hunter:2007},
    \texttt{astropy}~\citep{2013A&A...558A..33A},
    \texttt{aplpy}~\citep{2012ascl.soft08017R}
 }

\bibliography{bibliography,software}{}
\bibliographystyle{aasjournal}

\appendix
\section{Uncertainties in the Transition Column and Volume Densities}\label{app:uncertainties}
As {described} in Section~\ref{sec:observations}, we rejected {sky pixels for which polarization measurements were considered too contaminated by reference beam flux to be of use. However, it is possible that even small levels of contamination may affect our analysis, as variations} may alter the value of the HRO parameter by modifying the relative orientation angle. To estimate the {magnitude} of this effect, we follow the procedure described in \cite{1997ApJ...487..320N} and \cite{2019ApJ...872..187C}. Assuming an unknown polarization angle, but {uniform polarization fraction of $10\%$} in the reference beam, we can compute $\Delta \hat{E}_{\text{ref}}$: 
\begin{equation}
    \Delta \hat{E}_{\text{ref}} = \frac{1}{2} \arctan \left[ \frac{p_r w}{\left(p_m^2 - p_r^2 w^2\right)^{\frac{1}{2}}} \right]
\end{equation}
This value represents the largest polarization angle `error' possible {due to} reference beam contamination. (See Section~\ref{sec:observations} for symbol definitions{.}) After calculating $\Delta \hat{E}_{\text{ref}}$, we then modify all of our measured polarization angles with this maximum `error' to produce two extreme cases. One case is produced by increasing the nominal $\hat{E}$ value {for all sky positions} by $\Delta \hat{E}_{\text{ref}}$; the other by decreasing $\hat{E}$ by $\Delta \hat{E}_{\text{ref}}$. After reapplying the procedures from Section~\ref{sec:methods} using these {modified} scenario polarization angles, we find that the resulting $N_{\text{H}_2, \text{tr}} $ value {changes by less than} a factor of $\sim 1.25$.

As the uncertainty in the polarization angle measurement is not directly encapsulated in the HROs and the subsequent linear regression to obtain $N_{\text{H}_2, \text{tr}}$, we now consider this source of uncertainty. In terms of its effect on the analysis, this uncertainty will play a similar role to that of the reference beam contamination. In {that case,} the median $\Delta \hat{E}_{\text{ref}}$ `error' value is computed {to be $\approx19\degr$.} By comparison, the median uncertainty in the polarization angle for data used in this study {is only $\approx1.1\degr$.} As a result, we expect that any effect of the uncertainty in the polarization angle measurement {will} be subdominant to that of the reference beam contamination, and thus, minimally affect the resulting $N_{\text{H}_2, \text{tr}}$ value.

The existence of reference beam contamination also presents an issue when merging polarimetric observations from different instruments, as we do here (e.g., Figure~\ref{triple-figure}). Owing to differences in referencing strategies and methods, the role of the reference beam can vary greatly depending on the instrument. As noted above, {reference beam contamination} can alter the observed polarization angle{, which} can severely impact {HRO} results. Thus, a thorough investigation is required prior to any combined analysis. As the \planck instrument does not employ referencing methods in its observations, {\planck data} do not suffer from this type of contamination.

However, {other complications} may arise from conducting a combined analysis of \planck and \hawcplus polarimetry. The angular resolutions of these two datasets differ; the \planck data {have} a FWHM resolution of {$5\arcmin$ and} our \hawcplus data are at $36.3\arcsec$. As a result of their lower resolution, the \planck observations are unable to resolve any small-scale field disorder that may exist. A more disordered field would suggest less preferential alignment and thus a flatter HRO. The corresponding $\xi$ would potentially be closer to zero than nominally suggested by the \planck analysis. However, at the large scales and lower column densities investigated by \planck, the field is believed to be sub-Alfv\'enic~\citep{2016A&A...586A.138P}. The field is thus likely well-ordered suggesting that this effect is likely minimal.

An additional source of uncertainty is introduced by the relatively short observing wavelength of the \hawcplus polarimetry. At 154 $\micron$, the magnetic field traced by \hawcplus may not follow the column density structure mapped by \herschel at the submillimeter wavelengths, nor be directly comparable to the 850 $\micron$ (353 GHz) observations of \planck. It is possible that the shorter wavelength observations used here are less sensitive to the colder dust. Certainly, polarimetric observations at longer wavelengths would be beneficial. However, comparisons suggest that the variations between the measured magnetic field {orientation} at these wavelengths are typically not significant. Observations of OMC-1 at 53, 89, 154, and 214 $\micron$ using \hawcplus and at 850 $\micron$ using POL-2 show general agreement in magnetic field {orientation} despite the wavelength difference~\citep{2017ApJ...846..122P, 2019ApJ...872..187C}. This is also true when comparing the 850 $\micron$ POL-2 observations of \rhoopha from \cite{2018ApJ...859....4K} with the 154 $\micron$ observations presented in \cite{2019ApJ...882..113S} (and here). 

 {As stated in Section~\ref{subsec:results:transition}, we estimate that the total uncertainty resulting from the effects discussed here to be a factor of three on our value of  $N_{\text{H}_2, \text{tr}}$. However, these effects are subdominant compared to the sampling uncertainty (also described in Section~\ref{subsec:results:transition}).}

\end{document}